# Hysteresis and trap emission in dc-biased integrated lithium niobate electro-optic modulators

*Matthew Yeh[1], C. J. Xin[1], Donald Witt[1], David R. Barton III[1,a], Evelyn L. Hu[1,*], Marko Lončar[1,*]*

[1]John A. Paulson School of Engineering and Applied Sciences, Harvard University, Cambridge, MA 02138, USA.

[a]Present address: Department of Materials Science and Engineering and the Materials Research Center, Northwestern University, Evanston, IL 60208, USA.

*Corresponding author. E-mail: ehu@seas.harvard.edu, loncar@g.harvard.edu

**The electro-optic effect is crucially important for low power and efficient tuning of integrated photonic circuits. However, in electro-optic materials such as lithium niobate, dc biasing for an extended duration of time results in the emergence of numerous nonidealities, including hysteresis—a persistent degradation of the magnitude and linearity of the dc electro-optic response. We show that electro-optic hysteresis can be reversed under both zero-bias and reverse-bias conditions, given sufficient time or reverse voltage and consistent with a defect model of the underlying physics. Specifically, we find that drift phenomena at short time scales can be explained by charge trapping dynamics near the contact junction, and thereby devise an active reset protocol that restores the magnitude of the response and partially restores the drift time scales.**

Owing to its large Pockels coefficient (~30 pm/V), lithium niobate has achieved great success as the workhorse electro-optic (EO) material, forming the backbone of optical modulators used in long-haul communications[1–3]. The advent of thin-film lithium niobate (TFLN) has enabled active photonic circuits to become increasingly compact, promising low-loss, high-bandwidth, and efficient operation with new functionality on a wafer scale[4–6]. To utilize the entirety of the application space will require reconfigurability via dc bias, from the single modulator level to fully integrated systems. However, while EO is well-behaved at microwave frequencies, it is commonly observed in many Pockels materials that different phenomena dominate at dc and low frequencies (<1 MHz), resulting in drift of the dc bias point[7–13]. Here we investigate the electronic device physics of an EO material by studying the repeatability of the dc response, using TFLN as a prototypical example. Previous work on electro-optic hysteresis (also known as applied-voltage fatigue or non-volatile drift memory) in bulk LN modulators has seen a litany of qualitative behaviors, including irreversible drift and increased drift rates after continuous biasing[14,15]. We elaborate on these phenomena in TFLN, and further demonstrate a protocol based on our understanding of the defect physics to restore the linearity and magnitude of the dc response. These insights should improve the reliability of EO modulators in long-term use cases.

We begin by characterizing dc drift over several days. A schematic of the experimental setup for measuring dc EO response is shown in Figure 1a. Laser light is coupled into a coupled resonator device and the optical mode splitting is monitored after voltage application, to infer the induced phase shift and its time evolution (drift). We define EO response as the change in splitting, normalized to the applied voltage (30 V here). We use coupled resonators because splitting is a differential quantity, thereby rejecting common-mode noise. The optical layer is defined by e-beam lithography and Ar-ion etching, and the electrodes are deposited directly in contact with TFLN by Ti/Au evaporation and lift-off (Fig. 1a, inset). No thermal annealing is undertaken for the devices in the main text of this paper.

To illustrate the notion of EO hysteresis, we first present an oxide-cladded device. Consecutive measurements of the step response show that the dc EO behavior changes significantly after long-term biasing (Fig. 1b). Instead of reproducing the initial measurement, the drift curve qualitatively extends the initial decay as if the bias were continually applied, corresponding to a persistent degradation of the EO response magnitude. The initial multiexponential drift curve is often understood from a circuit model perspective, where lumped-element resistances and capacitances provide the RC time constants that model how the charge distribution evolves in time[7,8]. However, such a model cannot explain persistent phenomena, i.e. if discharging took a macroscopic amount of time, the index shift would persist even after the voltage is turned off. For our devices, this is not observed. Instead, we interpret our results from a defect physics perspective, where each defect process contributes a lifetime component[10,16]. In this interpretation, numerous persistent electronic phenomena could contribute to EO hysteresis, including long-lived trap states with a low emission rate in lieu of a bias, ionic diffusion, and partial domain inversion[9,17–19]. To this end, we studied the evolution of hysteresis over time and as a function of voltage to elucidate deeper understanding of the microscopic mechanisms.

For the remainder of this study, we employed a TFLN chip that underwent an optical packaging process using photonic wire-bonding[20] to facilitate long-term measurement. These devices additionally underwent oxide cladding and via routing steps to enable a push-pull electrode configuration[21] (Fig. 2a, right inset). The push-pull configuration increases the signal-to-noise ratio (SNR) of the measurement compared to the single-drive configuration in Figure 1 by increasing the amplitude of the EO response by a factor of 4.

We monitored the evolution of the EO response in the absence of a bias to elucidate hysteresis reversal solely driven by thermal processes (Fig. 2a). The device is initialized by dc biasing at 30 V for 36 h, driving the drift curve into a steady-state region and therefore saturating all defect processes. Then, the voltage dependence of the EO shift (EO-V) is sampled every hour and the change in optical mode splitting at 25 V is taken as a representative value (Fig. 2a, left inset). Surprisingly, the EO response does not monotonically recover, but instead features an enhancement and subsequent slow decay toward the native (initial) magnitude (i.e. Fig. 2b).

To gain insight into this feature, we also inspected the linearity of the EO-V curve. After the initialization pulse, the initial linear EO-V curve degraded by 4.5x in magnitude and became rectifying (Fig. 2b,c). This is characteristic of a diode response, from which we can infer the emergence of a potential barrier localized to the metal-LN contact interface, possibly due to charge trapping or ion migration adjacent to the electrode[22]. The disappearance of this barrier and the reemergence of linear EO-V (Fig. 2d,e) suggest the motion or emptying of defect states away from this interface. The spatially localized nature of the metal-LN junction could support an ion migration model, where the change of the EO-V curve is most simply described by the time evolution of a Gaussian diffusion solution at a fixed distance from the interface, reminiscent of the initial enhancement of the hysteresis recovery. In contrast, a bulk-limited model agnostic of the specific spatial distribution of defect states would simply follow a multi-exponential, reminiscent of the long-time decay of the recovery curve. We find that neither sufficiently describes the full evolution of the passive hysteresis recovery curve, indicating that there could be multiple mechanisms at play (Fig. S1).

In conjunction with EO-V measurements, the drift curve itself ultimately reflects the degree to which defect processes that contribute to hysteresis have been reversed. Although linearity may be restored in the EO-V, a quicker decay back into the saturated regime (relative to the native drift) indicates that the defect processes have only partially reversed (Fig. 2f, +12h or +1 week). The time constant of this decay increases as the wait time increases, until the native drift is fully reproduced at a wait time of 3 weeks.

Motivated by standard defect engineering techniques in semiconductor devices, we explored the possibility of actively enhancing the recovery rate. Specifically, the emergence of Schottky-like behavior suggests that a reverse bias could be used to widen the depletion region and empty charge traps[23]. We therefore characterized the voltage-dependence of the drift. We observe an exponential growth of the bias drift rate (Fig. 3a), extracted by taking the slope from linear fits to the fast drift (timescale ~s) immediately after applying voltage steps in a staircase waveform. Phenomenologically, this functional form can be understood by considering a model in which the EO response is determined by the junction capacitance and its evolution with time and applied bias voltage, $EO(t,V) \propto C_J(t,V)$. For example, this may occur in the limit of capacitive voltage division where the junction capacitance is smaller than any other effective capacitances in the series circuit representing the modulator. The junction capacitance depends on the trapped charge density $n_T(t,V)$

$$C_J(t,V) = C_{j0}\sqrt{1-\frac{n_T(t,V)}{N}} \approx C_{j0}\left(1-\frac{n_T(t,V)}{2N}\right) \tag{1}$$

which changes as charges are trapped or emitted by the defects, at a rate determined by the applied bias. $C_{j0}$ is the zero-bias junction capacitance, and $N$ is the doping density of the TFLN. In the low-trap density limit, this suggests the magnitude of the EO efficiency is linearly related to the trapped charge density. For example, this could be justified in the case of congruent LN, which has a native niobium antisite $Nb_{Li}$ concentration of ~1%[24]. The trapped charge density follows an exponential decay with time,

$$n_T(t,V) = n_{T0}e^{-e_T(V)t} \tag{2}$$

where for simplicity we have assumed a detrapping process with voltage-dependent rate $e_T(V)$. This suggests that for a given voltage, which fixes the (de)trapping rate, the evolution of $EO(t)$ at short times after applying a step voltage reflects the time-evolution of the (de)trapping processes in an EOM. In practice, it is difficult to make quantitative claims about microscopic trap densities as many of the relevant parameters, such as native doping density and Schottky barrier height, are ill-characterized in TFLN.

The voltage dependence can be understood as barrier lowering and/or thinning due to the application of the reverse bias. Larger voltage bends the bands more, allowing either tunneling through a thinned potential barrier or thermal emission over a lowered barrier (Fig. 3b). In general,

$$e_T(V) = e_T(0)e^{\frac{f(V)}{k_BT}} \tag{3}$$

where $f(V)$ reflects the voltage-dependent barrier lowering. Thus, the drift rate, defined as the time-derivative of the EO response, simply reflects $e_T(V)$:

$$\frac{dEO}{dt}(V) \propto \frac{d}{dt}C_j(t,V) \propto -\frac{d}{dt}n_T(t,V) = e_T(V)n_T(t,V) \tag{4}$$

In the short-time limit, we can linearize Eq. (2), which also drops the time dependence of the drift rate in (4):

$$\frac{dEO}{dt}(V) \propto n_{T0}e_T(V) = n_{T0}e_T(0)e^{\frac{f(V)}{k_BT}} \tag{5}$$

Thus, the voltage-dependence of the initial drift rate provides insight into the effect of reverse biasing on the interplay between defect processes and potential barrier modification in EO devices. We find that the experimental data is fit well using a Poole-Frenkel model of trap emission[25,26], where $f(V) \propto \sqrt{V}$ and the field-enhancement is due to increased thermal emission over a lowered barrier (Fig. 3b).

The exponential growth of the drift rate indicates that applying a large voltage of opposite sign to the initial burn-in polarity can exponentially speed up the removal or redistribution of defects that engender hysteresis. We thus implemented a protocol to exploit this. We initially employed a slow reverse voltage ramp from 0 to -60 V over 1 h, observing that both the linearity and magnitude of the dc EO response can be recovered to match the initial performance (Fig. 4,a-c). However, the corresponding drift response curve demonstrates that the defect processes have only partially been reset—the post-reset drift curve does not reproduce the native drift, instead incorporating a fast initial decay that quickly returns to the saturated drift response (Fig. 4d). Importantly, this indicates that the electronic mechanism underlying (non)linearity at short time scales (~s) only partially accounts for long-term drift (~h).

We consequently explored longer active reset times, employing a trapezoidal pulse (Fig. 4d). In this case, the drift curves dilate and the time constants become comparably long to the initial drift curve, but still do not recover the exact functional form of the initial drift. We surmise this may be due to the activation of new drift pathways due to the difficulty of calibrating the exact reverse bias time, or due to the higher voltages employed. For practical applications however, the recovery of efficiency and reduction of the drift rates allow active feedback systems to be deployed, within the specifications of a lockbox's output[27]. Better optimized pulse sequences could be developed using higher voltage and shorter pulses to quickly reset the system.

We have shown that dc bias hysteresis, although apparently irreversible if applying the same polarity, can be reversed thermally given sufficient time or actively at an accelerated rate using a sufficiently large bias of the opposite polarity. The electro-optic techniques presented here act as an optical probe of electronic properties such as charge trap densities and emission rates. Thus, in conjunction with purely electronic techniques such as deep-level transient spectroscopy to extract standard semiconductor device parameters, voltage-pulsed EO measurements could be a powerful tool to identify dominant defects and their position in the bandgap. Extending these studies to different processing conditions and contact treatments (Fig. S2) will also be a powerful tool to elucidate possibly different hysteresis mechanisms in electro-optic materials including and beyond TFLN, such as TFLT, BTO, and AlN[11–13,28].

## ASSOCIATED CONTENT

**Supporting Information:** Additional comparison of models of hysteresis recovery, and hysteresis under different electrode geometries.

## ACKNOWLEDGMENTS

**Funding:** This work is supported by the National Science Foundation (OSI-2137723, EEC-1941583) and the Defense Advanced Research Projects Agency (HR0011-24-2-0360). M.Y. acknowledges support from the Department of Defense (DoD) through the National Defense Science and Engineering Graduate (NDSEG) Fellowship Program. The research D.B. performed was supported by an appointment to the Intelligence Community Postdoctoral Research Fellowship Program at Harvard University, administered by Oak Ridge Institute for Science and Education through an interagency agreement between the US Department of Energy and the Office of the Director of National Intelligence. The views, opinions and/or findings expressed are those of the authors and should not be interpreted as representing the official views or policies of the Department of Defense or the US Government. Device fabrication was performed at the Harvard University Center for Nanoscale Systems (CNS), a member of the National Nanotechnology Coordinated Infrastructure Network (NNCI), which is supported by the National Science Foundation under NSF award no. 1541959.

**Author contributions:** M.Y. conceived the idea, conducted the experiments, analyzed the data, and wrote the manuscript. C.X., D.W., and D.R.B. fabricated devices. E.L.H. and M.L. supervised the project. All authors discussed the results and commented on the manuscript. We thank Benjamin Fortuin for helpful discussions.

**Competing financial interests:** M.L. and C.X. are involved in developing lithium niobate technologies at HyperLight Corporation.

**Data and materials availability:** The materials that support the findings of this study are available from the corresponding author upon reasonable request.

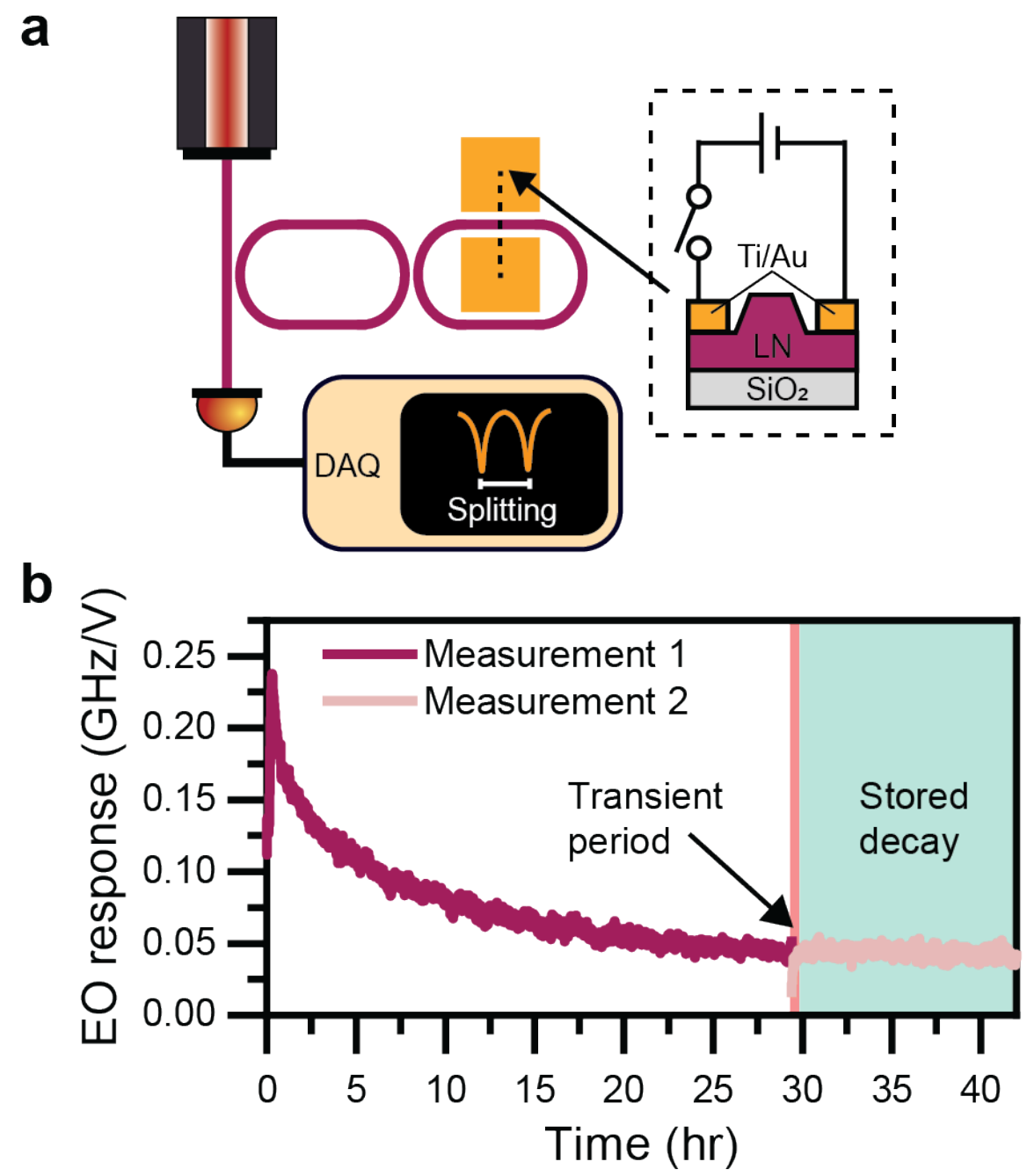

**Figure 1. Measurement and concept of electro-optic hysteresis. (a)** Experimental setup for measuring dc bias phenomena in electro-optic devices. Voltage is applied through a probe and the change in optical mode splitting is monitored over time. (Inset) Device cross-section, showing direct metallization in contact with the LN. **(b)** Consecutive dc drift measurements for an oxide-cladded TFLN modulator (unannealed), showing electro-optic hysteresis as a continuation of the drift decay, after a short transient period. This device was measured in a single-drive configuration, as depicted in (a).

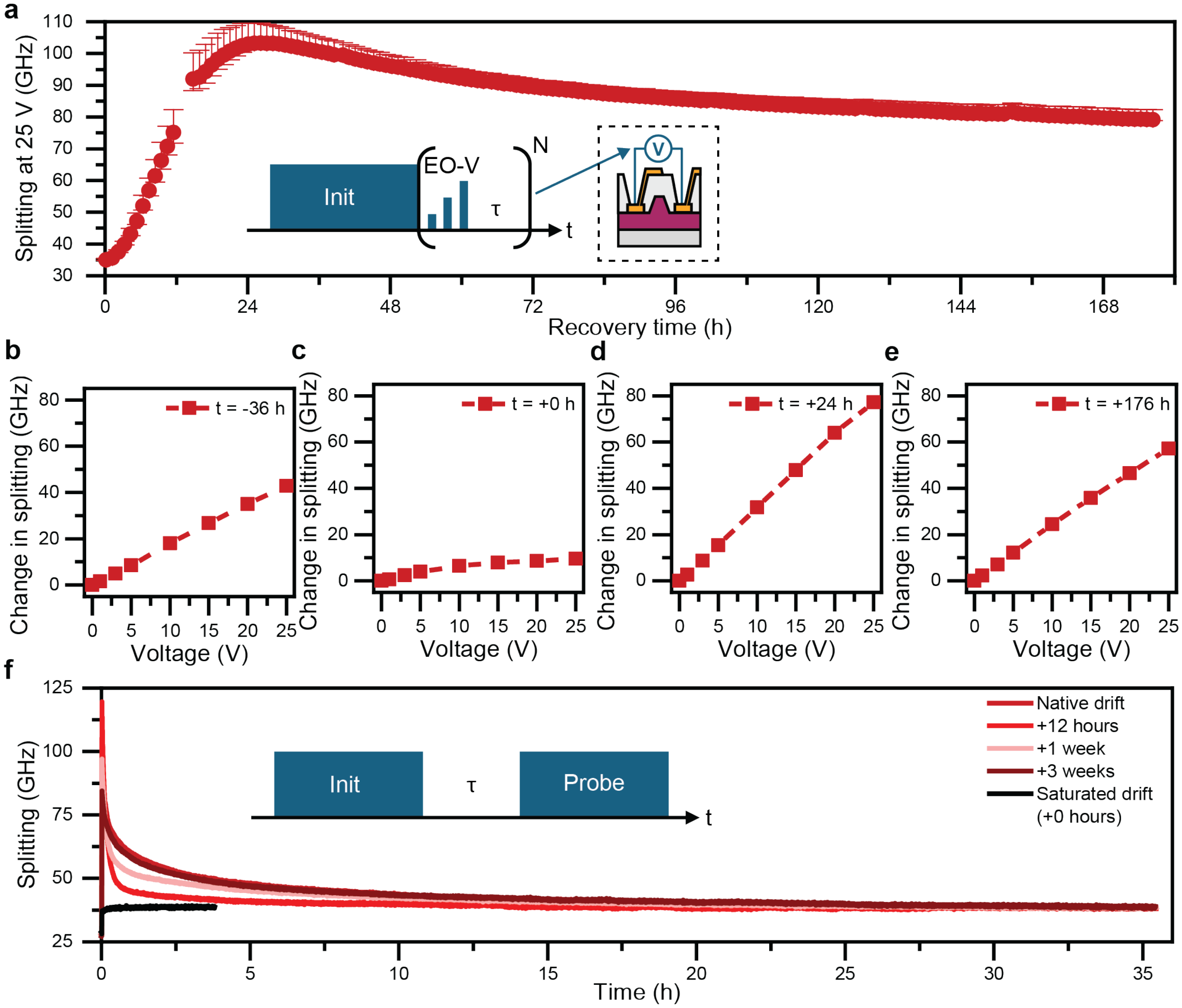

**Figure 2. Passive hysteresis recovery. (a)** Evolution of the EO response over the course of one week. The EO-V is sampled periodically and the frequency shift at 25 V is taken as a characteristic value of the EO response. The frequency shift is extracted by taking the mean value over 1 min and error bars indicate the 10th and 90th percentiles of this time series. (Inset) Measurement sequence (left) applied to the device (right), showing an initial 36 h initialization measurement to saturate the response, before repeatedly measuring EO-V. Cladding oxide was deposited using a low-temperature CVD process to facilitate a via process used to operate the coupled resonators in a push-pull configuration, increasing EO response (and SNR). **(b-e)** EO-V characteristics (b) prior to measurement initialization (t=-36 h), and after rest times of (c) 0 h, (d) 24 h, and (e) 176 h, showing the evolution of the voltage dependence from rectification to temporary enhancement. The optical mode splitting at zero bias is ~28 GHz. **(f)** Initial dc drift curve and drift curves after different recovery times, showing continued saturation of the drift immediately after the initial measurement. The "+3 weeks" curve lies on top of the "Native drift" curve, indicating complete recovery. (Inset) Measurement sequence. Each drift curve is acquired after a separate initialization.

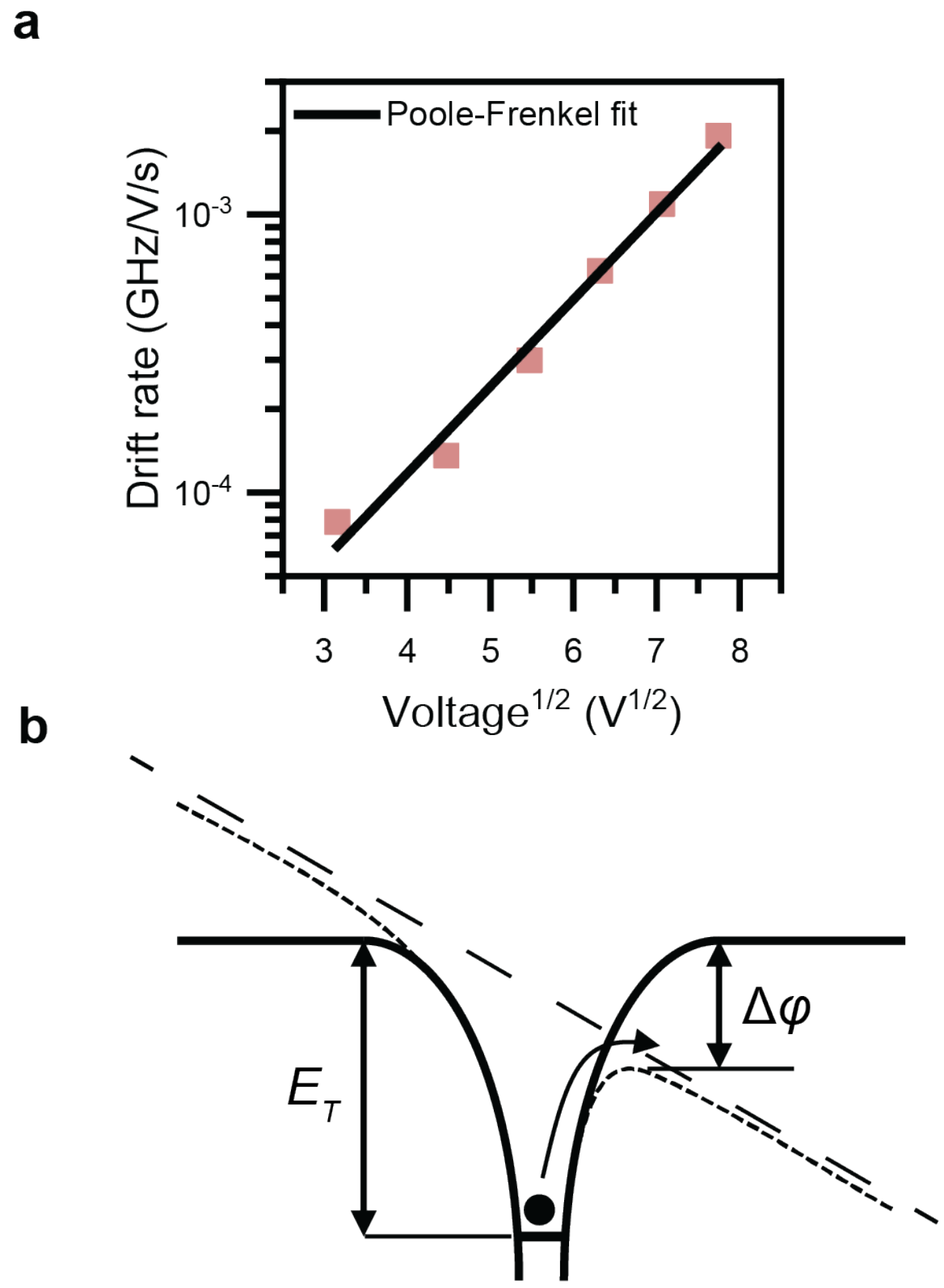

**Figure 3. Field-enhanced defect emission. (a)** Reverse voltage dependence of drift rate, showing an exponential growth with the square root of field. **(b)** Energy diagram of the field enhancement process. Traps thermally ionize from a well of depth $E_T$ with a characteristic decay rate that is voltage-dependent, owing to the lowering of the barrier by $\Delta\varphi$ with increased reverse bias.

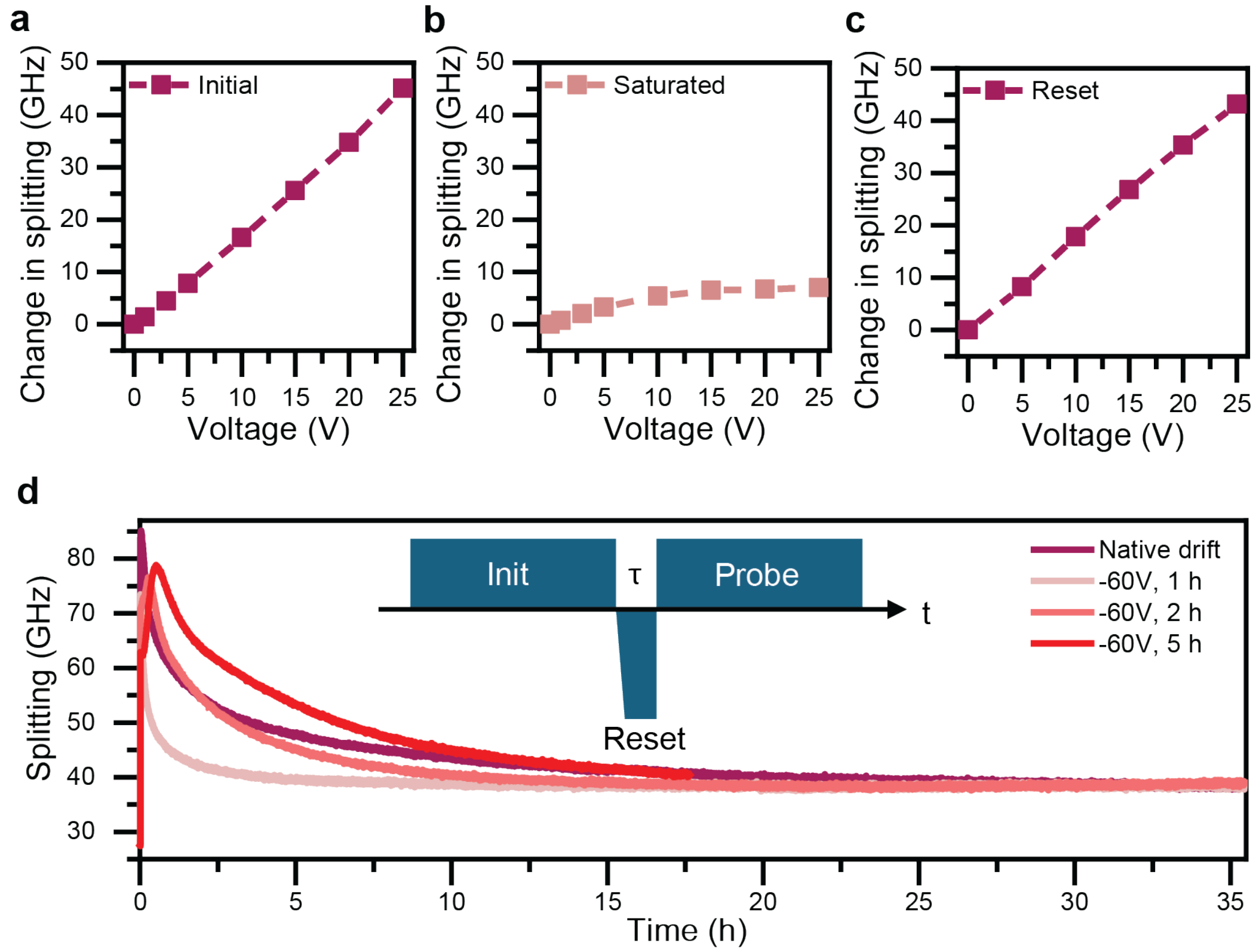

**Figure 4. Voltage-induced hysteresis recovery. (a-c)** EO-V characteristics (a) before and (b) after the initialization measurement, and (c) after applying -60 V over the course of 1 h, showing recovery of the initial linear curve. **(d)** Initial dc drift curve and dc drift curves after different reverse bias times. Qualitatively, increasing the reverse bias time dilates the drift curve, but does not fully recover the native drift despite the recovery of linearity, indicating the presence of different underlying electronic mechanisms. (Inset) Measurement sequence including a trapezoidal (ramp + hold) reset pulse of opposite polarity.